\documentclass[twocolumn,aps,pra,longbibliography]{revtex4-2}
\usepackage{amsmath}
\usepackage{amssymb}
\usepackage{graphicx}
\usepackage[english]{babel}
\usepackage[utf8]{inputenc}
\usepackage[T1]{fontenc}
\usepackage{color}

\usepackage{braket}
\usepackage{soul}

\begin{document}

\title{Chaos and Diffusion in Twisted Bilayer Graphene}

\author{Olga Arroyo-Gascón}
\address{Universidad de Salamanca IUFFyM, E-37008 Salamanca, Spain}
\thanks{Present address: Institut de Physique Théorique, Université Paris-Saclay, CEA, CNRS, F-91191 Gif-sur-Yvette, France}

\author{Manuel Pino}
\address{ Universidad de Salamanca IUFFyM, E-37008 Salamanca, Spain}

\date{\today}

\begin{abstract}

We numerically analyze chaos and diffusion in experimentally-relevant models of twisted bilayer graphene. Our results indicate that finite systems at both commensurate and incommensurate rotation angles exhibit chaos. We compute the Thouless energy of the systems at intermediate twist angles to further understand the diffusive processes associated to the chaotic nature of the spectrum. We find that the mean free path of the non-interacting electrons scales with the system size, which is consistent with diffusive processes caused by border-induced scattering. In summary, our results show the role of borders as a experimentally relevant single-particle scattering mechanism, which is much stronger than in single-layer graphene and induces chaos and diffusion regardless of the rotation angle and the lattice commensurability.\end{abstract}
\maketitle

\section{Introduction}
\label{sec:intro}

Twisted and layered structures have recently been the focus of theoretical and experimental breakthroughs in quantum materials. One of the most paradigmatic examples is twisted bilayer graphene (TBG), which at a small relative rotation angle hosts, among others, strong electronic correlations, topological phases and unconventional superconductivity \cite{Cao20181,Cao20182,Andrei2021}. The emergence of these phenomena has motivated the search for other twisted materials, mostly including graphene-based systems \cite{Zhang2021,Cao2020,ArroyoGascn2020,ArroyoGascn2023,Chen2019} and transition metal dichalcogenides \cite{Regan2020,Shimazaki2020,Wang2020}, and even coined a new field, twistronics \cite{twistronics}.

All these materials share the appearance of moiré patterns that arise when superimposing two or more rotated layers. This procedure can significantly alter their properties, and induces a different  
periodicity than the one that characterizes each separate layer. The majority of works on this topic have successfully characterized layered materials in the crystalline limit, where the system has translational symmetry and the Bloch theorem applies \cite{bistritzer_moire_2011}. However, there are important questions which cannot be answered within this framework, such as the emergence of chaos and diffusion. These processes can arise due to scattering of Bloch states, so that they mix and decay, or due to a complex unit cell\ \cite{porter_chaos_2017}.  The specific scattering mechanisms characterize experimentally measurable properties of layered materials, as its conductivity \ \cite{clerico_quantum_2019, estrada2025superballistic,zhang2026imaging} or photo-response\ \cite{deng2020strong,delgado2025unveiling}. An understanding of those mechanisms in twisted layered materials is therefore of high relevance.

There are several plausible processes that can give rise to chaos and diffusion in TBG. The most obvious 
sources are electron-electron and phonon-electron interactions, or impurity scattering 
\ \cite{hwang2020impurity, birkbeck2025quantum,ray2016electron}. Previous works have recently explored moiré single-particle models of incommensurate TBG lattices. Such a scenario has been approached in $30^\circ$-rotated TBG, which is a dodecagonal quasicrystal  \cite{uri_superconductivity_2023,yu_dodecagonal_2019,yu_electronic_2020,park_emergent_2019,vidarte_quasicrystalline_2024,yao_quasicrystalline_2018}, frequently by resorting to commensurate approximants. In fact, quasicrystalline layered graphene has already been produced experimentally \cite{uri_superconductivity_2023}. Besides $30^\circ$, the electronic structure of incommensurate bilayer graphene has been studied in Refs. \cite{chalin_tight-binding_2022,koshino_electronic_2015,koshino_interlayer_2015}, but mainly with the aim of performing band structure calculations. 

Ref. \cite{goncalves_incommensurability-induced_2022} focuses on TBG at different rotation angles under twisted boundary conditions, and computes signatures of chaos in the spectrum and wavefunctions. The authors observe chaotic behavior in the subgap states near the magic angle $\theta \sim 1.1^\circ$ only for incommensurate structures, showing that commensurability can have a drastic effect on the chaotic properties of TBG as well as on its conductivity. Recently, TBG cavities where also found to exhibit chaos due to a a dominant trigonal-warping mechanism, although edge effects where found to become more relevant as a source of chaos at increasing rotation angles\ \cite{lin2026shaping}.

Here, we numerically probe quantum chaos in finite-sized TBG flakes. Additionally, we compute the Thouless energy, which conveys valuable information about the diffusive behavior of the system. Although this energy has been recently computed in graphene bilayers in the context of Josephson junctions \cite{Hu2023,Xie2023}, this work provides the first Thouless energy analysis in the context of chaotic dynamics in TBG, to the best of our knowledge.

In order to prove these points, we first introduce the employed model in Sec.\ \ref{sec:model}, and later provide evidences of the chaotic nature of TBG flakes at any rotation angle in Sec.\ \ref{sec:quantumchaos}. The final section, Sec.\ \ref{sec:ethouless}, is devoted to explaining the computation of the Thouless energy in our system and to analyzing its scaling with the system size.

\section{System geometry and methodology}
\label{sec:model}
The moiré patterns exhibited by TBG geometrically induce a new periodicity for the bilayer rotated system. We denote the reciprocal lattice vectors of each layer by $\mathbf{b}_i$ and $\tilde{ \mathbf{b}}_i=R_\theta \mathbf{b}_i$, where $i=1,2$ and $R_\theta$ is a rotation transformation. Then, the periodicity of the moiré pattern in reciprocal space will be described by two new vectors $\mathbf{g}_{Mi}=\mathbf{b}_i-\tilde{ \mathbf{b}}_i$ \cite{koshino_electronic_2015}. This will yield a real-space periodicity 
$L_M=2\pi/|\mathbf{g}_{Mi}|=a/[2\sin(\theta/2)]$, where $a$ is the lattice constant of graphene 
\cite{Cao20181,koshino_electronic_2015,moon_optical_2013}. This moiré periodicity gives rise, for instance, to visually recognizable triangular patterns. Note that this moiré periodicity does not necessarily imply a crystalline lattice, which is a more stringent condition as it should meet commensurability in that case.

TBG forms a commensurate lattice if the rotation between layers $\theta$ maps a vector $n\mathbf{a}_1+m\mathbf{a}_2$ to a vector $m\mathbf{a}_1+n\mathbf{a}_2$, where $n$ and $m$ are integers and $\mathbf{a}_1$, $\mathbf{a}_2$ are the real-space lattice vectors of a graphene sheet\ \cite{koshino_electronic_2015,moon_optical_2013,suarez_morell_flat_2010}.  Although  the rotation angle $\theta$ can span a continuous range of values, only a discrete number of them will give rise to a commensurate unit cell. For these commensurate angles, dubbed $\theta_{com}$, we find the following condition \cite{mele2010,TramblydeLaissardire2010}:
\begin{equation}
\cos{\theta_{com}=\frac{n^2+m^2+4nm}{2(n^2+m^2+mn)}}.
\label{eq:commangle}
\end{equation} 
It is also possible to extract  the lattice constant of the resulting superlattice unit cell as $L= |m-n|L_M$ \cite{mele2010,TramblydeLaissardire2010}, which is a multiple of the general-case moiré periodicity $L_M$ defined in the previous paragraph.

\begin{figure}[h!]
    \centering
    \includegraphics[width=0.4\textwidth,trim={3cm 8cm 20cm 0cm},clip]{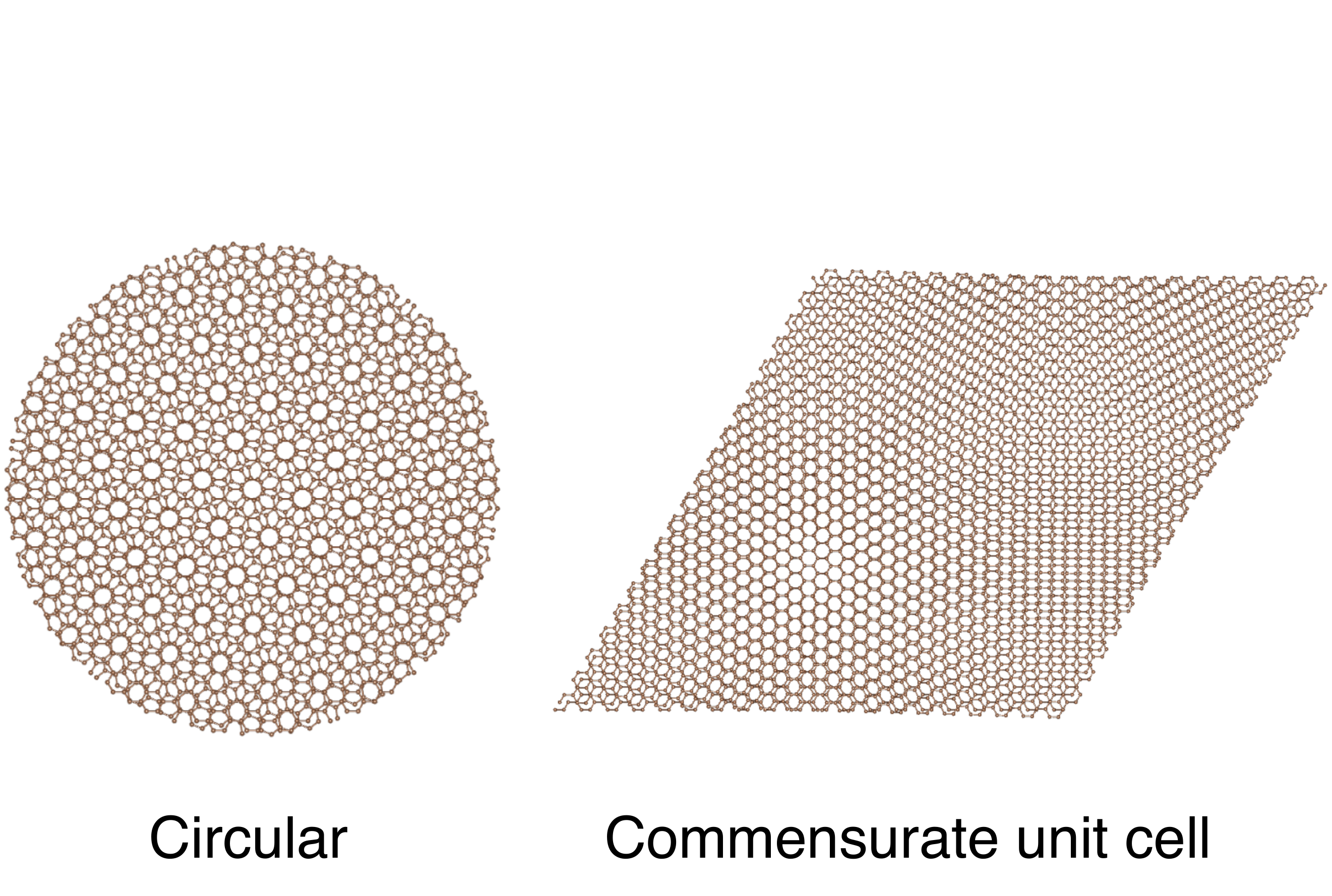}
    \caption{The two different types of TBG flakes studied in this work. The circular flake corresponds to a rotation angle $\theta_{com} = 21.8^\circ$, and the commensurate unit cell flake to $\theta_{com} = 1.05^\circ$. }
    \label{fig:tipos_geometria}
\end{figure}

In this work,  we construct TBG lattices under different geometries, as depicted in Fig. \ref{fig:tipos_geometria}. First, circular TBG flakes are constructed under open boundary conditions using the NetworkX Python package \cite{networkx}. 
The graphene layers constituting these flakes are rotated an angle $\theta$, which can be incommensurate or
commensurate. The flakes with circular boundaries are too regular to induce chaos by themselves, as a circular (smooth) billiard is known to be described by integrable dynamics\ \cite{zhang2026Experimtnal}.
Finally, in order to create commensurate supercells, a TBG unit cell is generated by specifying their $n,m$ indices following the methodology of Refs. \cite{suarez_morell_flat_2010,SurezMorell2011}. Although in Fig. \ref{fig:tipos_geometria} we show a single commensurate unit cell, throughout our calculations we consider large supercells, in order to obtain a translationally-invariant system in the thermodynamic limit.  
These commensurate supercells have the shape of
a $60^\circ$ rational rhombus; 
rational rhombus billiards are known to be not chaotic, but pseudo-integrable\ \cite{Prozen_Quantum2021,Bogomolnymodels1999,gremaud1998spacing}.

As discussed in Section I, commensurate approximants are common in the theoretical modeling of TBG structures. For instance, Ref. \cite{goncalves_incommensurability-induced_2022} implements twisted boundary conditions and performs a scaling analysis based on two parameters, related to an approximant and to the system size, respectively. In our work, we avoid the use of approximants and work with open boundary conditions.

Once that we have reviewed the geometry of our TBG lattices, we resort to a tight-binding model to study their electronic structure, as it is an ubiquitous approach among TBG works 
\cite{koshino_electronic_2015,koshino_interlayer_2015,yao_quasicrystalline_2018,yu_dodecagonal_2019,chalin_tight-binding_2022,yu_electronic_2020,suarez_morell_flat_2010,twistronics,fang_electronic_2016,nam_lattice_2017}. We follow the tight-binding model introduced in Ref. \cite{nam_lattice_2017}, which considers inter- and intralayer interactions for each atom in the system, up to a certain cutoff distance. The Hamiltonian is
\begin{equation}
H= -\sum_{i,j} t({\mathbf R}_i - {\mathbf R}_j)  \ket{{\mathbf R}_i}   \bra{ {\mathbf R}_j } + \mathrm{h.c.},
\label{eq:TBHam}
\end{equation}
where $i$ and $j$ label the atomic positions within the two TBG graphene sheets.
The hopping parameter is
\begin{equation}
-t({\mathbf R}) = V_{pp\pi} (R)  \left[ 1-\left( \frac{ {\mathbf R} \cdot  \mathbf{e}_z } {R} \right)^2 \right] +  V_{pp\sigma} (R) \left( \frac{ {\mathbf R} \cdot  \mathbf{e}_z } {R} \right)^2.
\end{equation}
Here, $|\mathbf R|=|\mathbf R_i-\mathbf R_j|$ is the distance between atoms $i$ and $j$, and $\mathbf{e}_z$ is the unit vector in the out of plane z-direction. 

The $V_{pp\pi} (R)$ and $V_{pp\sigma} (R)$ terms are, explicitly,
\begin{align}
V_{pp\pi} (R) =  V^0_{pp\pi} \exp \left( -\frac{R - a_0}{ r_0} \right),\notag \\
V_{pp\sigma} (R) =  V^0_{pp\sigma} \exp \left( -\frac{R - d_0}{ r_0} \right).
\end{align}
Here, the nearest-neighbor distance is $a_0\simeq 1.42$ \AA, $r_0= 0.184\sqrt{3}a_0$ is a decay length, and $d_0= 3.35$ \AA\ is the interlayer distance between graphene layers \cite{nam_lattice_2017}. These parameters will be used throughout all calculations unless otherwise stated.

\begin{figure*}[t!]
    \centering
    \includegraphics[width=0.85\textwidth,trim={0cm 3cm 0cm 0cm},clip]{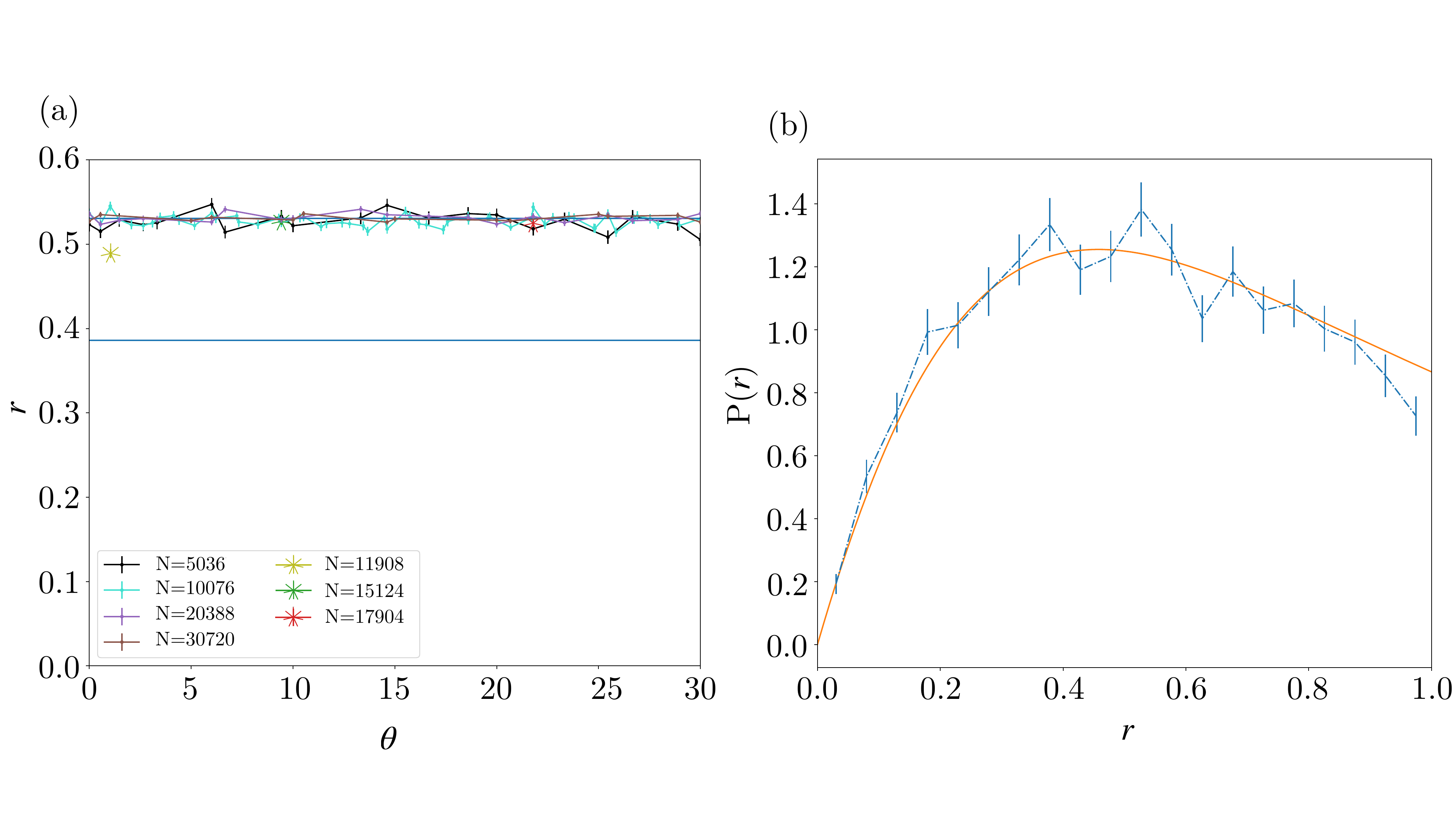}
    \caption{(a) Level statistics as a function of rotation angle $\theta$ for circular flakes (lines) and commensurate supercell flakes (stars), with diverse number of atoms $N$. For each system size, averages over all eigenstates within 3 and 5 eV are performed. 
    (b) Probability distribution of the level statistics for a commensurate supercell with $\theta_{com}=21.8^\circ$ (blue), and GOE probability distribution (orange). Averages over all
    eigenstates within 3 and 5 eV are performed.}
    \label{fig:rs_angle_p(r)}
\end{figure*}

In the following, we will analyze the presence of chaos in several TBG geometries and at different commensurate and incommensurate angles. First, we diagonalize the above Hamiltonian using SLEPC libraries\ \cite{hernandez2005slepc}. Then, we characterize its spectrum via quantities such as the adjacent gap ratios of density-density correlations between eigenstates. Notice that we are not averaging over different realizations of the previous Hamiltonian as it does not contain any random quantity. We will focus on an intermediate energy range, around $4$ eV, where 
the density of states is high and averaging is more robust. 

The presence of additional symmetries in a certain system can enforce degeneracies and consequently alter the energy spectrum, hindering the identification of chaotic properties. 
Depending on the rotation angle, commensurate TBG and TBG quasicrystals can host a large number of crystalline symmetries \cite{Yu2024,yu_dodecagonal_2019,yu_electronic_2020,park_emergent_2019,vidarte_quasicrystalline_2024,yao_quasicrystalline_2018}. From visual inspection of Fig. \ref{fig:tipos_geometria}, circular flakes show different mirror planes that may vary depending on $\theta$, and may host more crystalline symmetries than their commensurate unit cell counterparts. Therefore, in order to facilitate a more direct comparison between geometries and to construct a more realistic lattice, a random sliding between layers is introduced for all cases unless otherwise stated. 

\section{Quantum Chaos in twisted bilayer graphene flakes}
\label{sec:quantumchaos}

Using the tight-binding Hamiltonian of Eq.\ \ref{eq:TBHam}, we study the chaotic properties of the energy spectrum of several TBG lattices. As our Hamiltonian is time-reversal symmetric, we will rely on results from the Gaussian Orthonormal Ensemble (GOE) of Random Matrix Theory\ \cite{wigner1967random,Bohigas1984, mehta2004random}. We use  the adjacent gap ratio  $r_n=\min(s_n,s_{n-1})/\max(s_n,s_{n-1}),$ where $s_n=E_{n+1}-E_n$ is the nearest-neighbor spacing \cite{oganesyan2007localization}. In chaotic quantum systems, the local properties of the spectrum are dominated by level repulsion, which induces $\langle r_{GOE} \rangle \approx 0.536$ for the GOE ensemble\ \cite{Atas2013,Giraud2022} and a full distribution probability 
\begin{equation}
P_{GOE}(r)=7.75(r+r^2)/(1+r+r^2)^{5/2}.\label{eq:goefull}
\end{equation}
For an integrable system, the distribution of levels follows a Poisson law, which implies $\langle r_{P} \rangle \approx 0.386$ and a distribution probability 
\begin{equation}
P_{P}(r)=2/(1+r)^2 .
\end{equation}

Fig. \ref{fig:rs_angle_p(r)} (a) shows $r$ as a function of $\theta$ for circular flakes and commensurate supercells under open boundary conditions, up to $N\sim 31000$ atoms. Averages are computed over all states between 3 and 5 eV. We emphasize that 
a random sliding between the two layers is included in all cases (without averaging over different slidings). For the commensurate supercells, we have constructed lattices labeled by the unit cell indices $(n,m)$, introduced in Section II. Namely, we choose $(n,m)=$$(2,1)$, $(8,7)$ and $(32,31)$. From Eq. \ref{eq:commangle}, we find the corresponding commensurate rotation angles $\theta_{com} = 21.8^\circ$, $4.4^\circ$ and $1.05^\circ$, respectively, with a moiré superlattice periodicity ranging from $L_M=6.5$ \AA\ to $134$ \AA. 

From Fig. \ref{fig:rs_angle_p(r)} (a), we find that both circular and commensurate supercell flakes show a chaotic spectrum with $\langle r \rangle$ close to GOE, independently of the 
rotation angle. The only angle that shows a small deviation from GOE is $\theta_{com} =1.1^\circ$ (yellow star in the figure), where $r$ is slightly smaller than predicted. The chaotic behavior of the commensurate structures is further checked in Fig. \ref{fig:rs_angle_p(r)} (b), displaying in blue the full probability distribution $P(r)$ for the $(n,m)=(2,1)$ commensurate superlattice with $\theta_{com}=21.8^\circ$. This case is marked as a red star in panel (a) of the same figure. The results of Fig. \ref{fig:rs_angle_p(r)} (b) are again fully compatible with a GOE behavior given by the aforementioned $P_{GOE}(r)$ in Eq.\ \ref{eq:goefull}, and represented in orange in Fig. \ref{fig:rs_angle_p(r)} (b). 
We therefore do not find pseudo-integrable statistics as 
in Ref. \cite{gremaud1998spacing} for the geometries and energy ranges we focus on in this work.    Surprisingly, there is no notable difference in $r$ between incommensurate and commensurate rotation angles, except from the small deviation at $\theta_{com}=1.05^\circ$ which is still far from a Poisson law.

\begin{figure}[t!]
    \centering
    \includegraphics[width=0.5\textwidth,trim={6cm 6cm 24cm 0cm},clip]{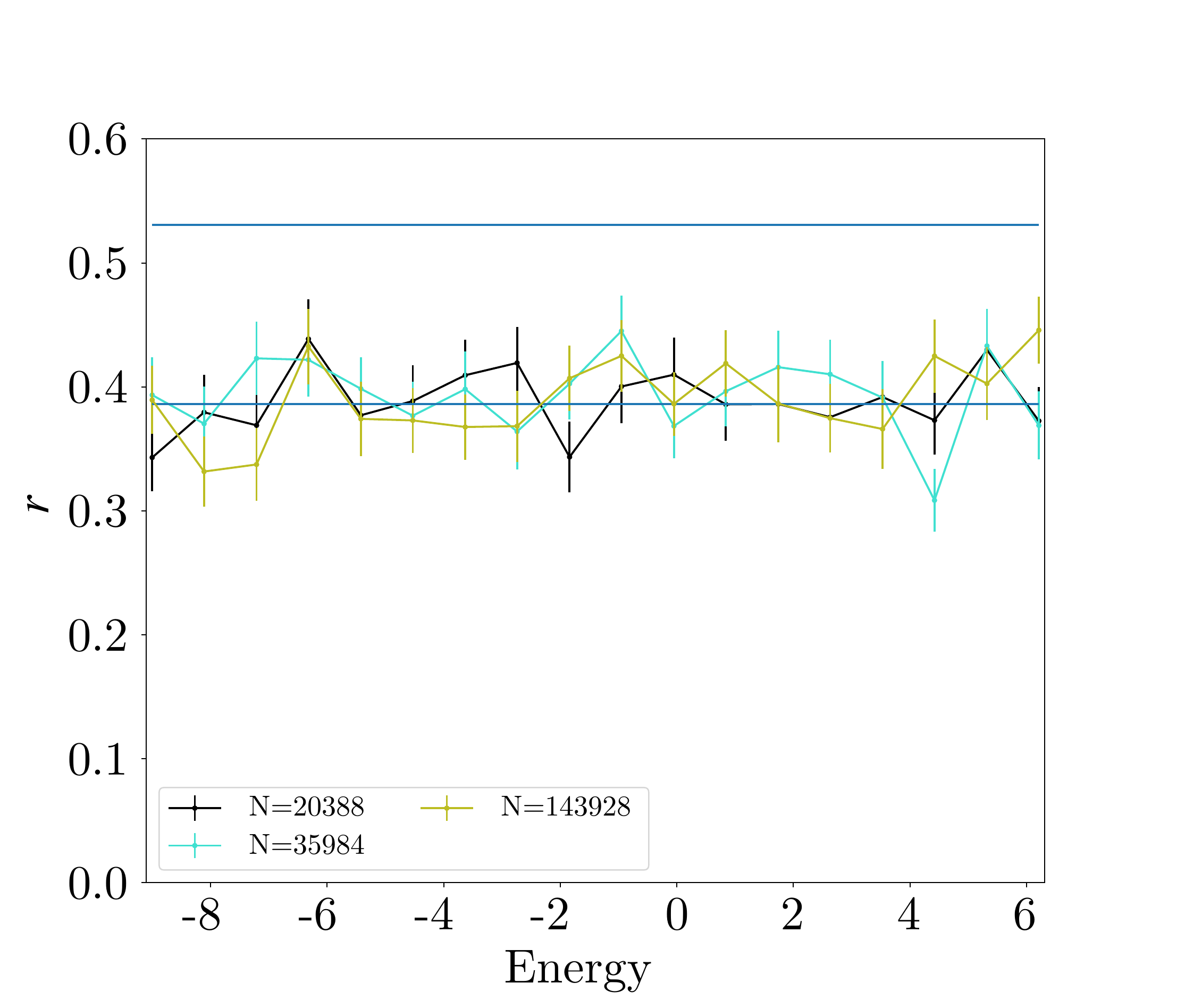}
    \caption{Level statistics as a function of energy for circular flakes at $\theta=0^\circ$ with diverse number of atoms $N$. For each system size, averages over 100 eigenstates within the energy range shown in the Figure are performed.}
    \label{fig:circ_0deg}
\end{figure}

To better 
identify the main contribution to chaos in our system, especially for commensurate structures, in Fig. \ref{fig:circ_0deg} we have computed the adjacent gap ratio $r $ for highly-symmetric ($\theta=0^\circ$) circular flakes without random sliding, up to $N\sim 144000$ atoms and as a function of the energy. These flakes are AA-stacked, so one layer is exactly on top of each other, resembling a single layer of graphene. Indeed, as one could expect in graphene, in Fig. \ref{fig:circ_0deg} we find an integrable non-chaotic behavior. This is interestingly in opposition to former Fig. \ref{fig:rs_angle_p(r)} (a), where chaos appears for the same system at $\theta=0^\circ$ but having added a random sliding between layers. 
 We note that $0^\circ$-TBG is always commensurate in the thermodynamic limit regardless of the sliding introduced between the two flakes. 
Thus, incommensurability does not seem to be the main factor that produces chaos in TBG flakes.

So far, we have demonstrated that quantum chaos is easily achieved in TBG, even for commensurate flakes.  The semi-empirical parameter values used throughout this work are those discussed in Section II and provided in Ref. \cite{nam_lattice_2017}, and are expected to yield the best agreement with experiments. For those values, as discussed throughout this Section, TBG flakes display a chaotic behavior. The presence of quantum chaos has been in fact reported in other graphene-based lattices such as nanoflakes and billiards \cite{rycerz_random_2012,rycerz_strain-induced_2013,libisch_graphene_2009,yu_gaussian_2016,Bao2015,Ponomarenko2008}, and in incommensurate TBG \cite{goncalves_incommensurability-induced_2022,lin2026shaping}. However, 
the aforementioned works select
an energy window very close to the Dirac point whereas we focus on higher energies.

Therefore,  TBG is likely to display chaos independently of the system size in the range of energies that we have studied. 
We identify two leading sources of chaos in our system. 
First, the complexity of the flake borders can produce non-regular scattering, giving rise to chaotic dynamics. Second, the complexity of the moiré unit cell can also originate chaotic dynamics, yielding
level repulsions of eigenvalues with different band index\ \cite{porter_chaos_2017}. This second mechanism 
could be the dominant source of chaos in the commensurate supercells of Fig.\ \ref{fig:rs_angle_p(r)} (a) with small rotation angles. This follows from Section II, where the real-space periodicity of the lattice was analytically described as a function of $1/\sin(\theta/2)$. For small $\theta$, the moiré unit cell can thus be rather large. If it is of the order of the system size, we would be in a regime where the energies of nearest-neighbors correspond to different band indices. In this setting, chaos can arise due to the complex dynamics in the moiré unit cell \ \cite{porter_chaos_2017}.
However, this cannot be the only source of chaos for large commensurate rotation angles, as in this case the system size is significantly larger than the moiré unit cell. Level repulsion, characteristic of quantum chaos, can then only be due to scattering and mixing of different Bloch vectors. 

In the following section, we will  fully clarify the role of borders and explain the main scattering mechanisms to induce diffusion in our systems at large rotation angles. 
Moreover, most theoretical efforts have been directed towards magic angle TBG, whereas larger rotation angles remain mostly unexplored from a quantum chaos perspective.

\section{Diffusion and Thouless Energy}
\label{sec:ethouless}

In the previous section, we analyzed the presence of quantum chaos by studying the spectrum of several TBG systems. We now focus on the Thouless energy $E_{Th}$, a fundamental quantity that relates quantum chaotic motion with a diffusive behavior\ \cite{edwards_numerical_1972,abrahams_scaling_1979}. This energy equals the width of the spectrum for which eigenstates show equal correlations than those of the relevant Random Matrix Ensemble. As such, it has played an important role in the understanding of a variety of phenomena, as the eigenstate thermalization hypothesis\ \cite{dalessio_quantum_2016}, metal-insulator transitions in disordered models\ \cite{abrahams_scaling_1979} or in many-body localized ones\ \cite{pino_multifractal_2017}. 

We are interested in relating this energy with the diffusive process of a particle in TBG. This can be achieved via the relation $E_{Th}=\hbar/\tau, $ with diffusion time $\tau= L^2/D$ for a system of linear size $L$ and diffusion constant $D.$ For a particle moving with a Brownian motion, the diffusion constant can be computed as $D= v l/2$,  where $v$ is the average velocity of the particle and $l$ its mean free path. The Thouless energy becomes then
\begin{align}\label{Eq:Eth}
    E_{Th} = \frac{\hbar v l}{2L^2}.
\end{align}
If the system is intrinsically chaotic, the mean free path is finite and the Thouless energy scales as $E_{Th} \sim 1/L^2.$ 
This is for instance the case 
for diffusion 
due to scattering with impurities, electron-electron interactions or phonon-electron coupling. On the other hand, for a system in which boundary conditions introduce chaos, the mean free path is $l\sim L$ so that $E_{Th} \sim 1/L.$ We 
will make use of these differences in the scaling of the Thouless energy to 
distinguish between boundary-induced or intrinsic chaotic behavior in TBG. 
The immediate first step is to describe the calculation of 
the Thouless energy in our system. 

\subsection{Numerical computation of the Thouless energy}

\begin{figure}[t!]
    \centering
    \includegraphics[width=\columnwidth]{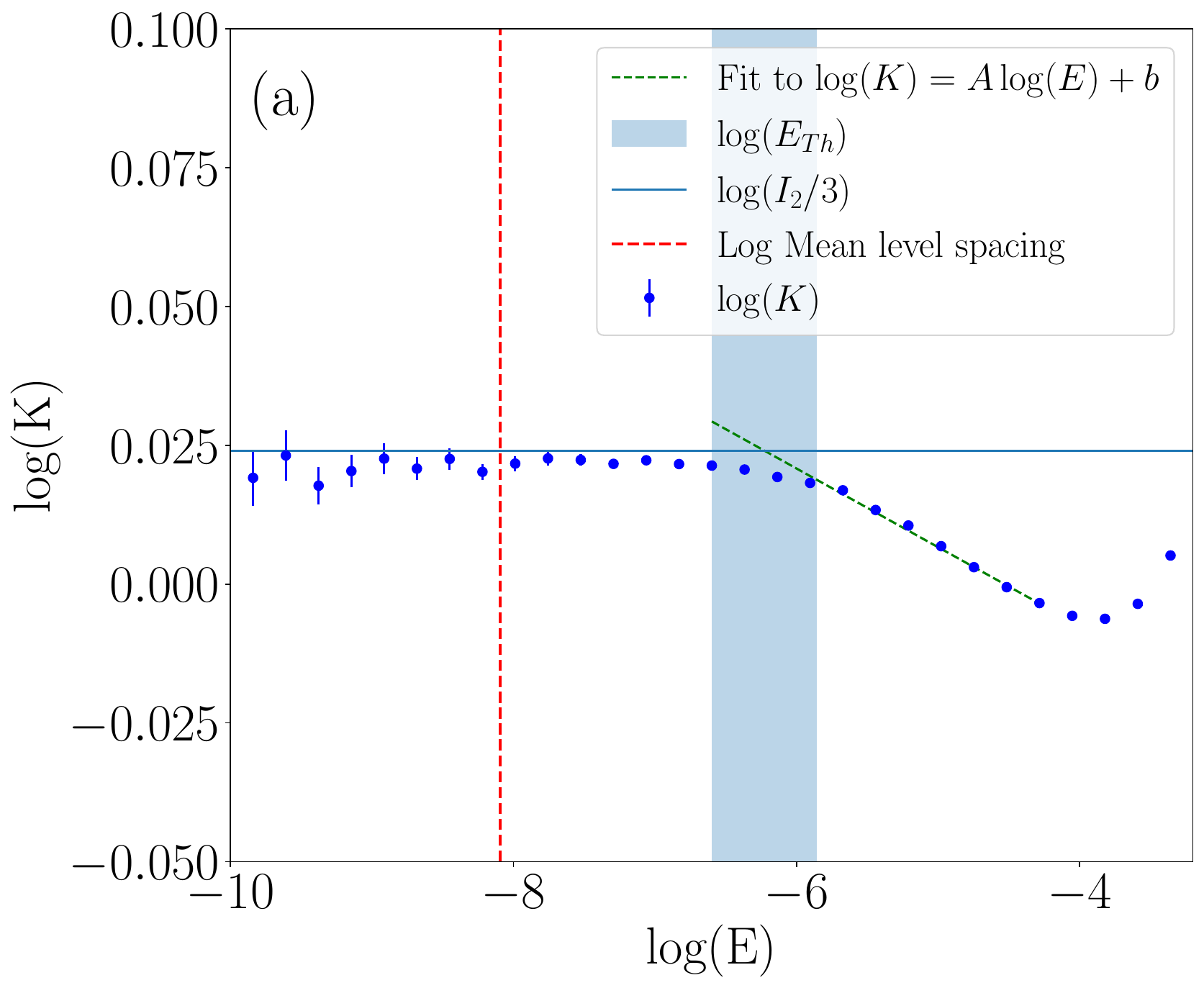}
    \vspace{0.5cm}
    \includegraphics[width=\columnwidth]{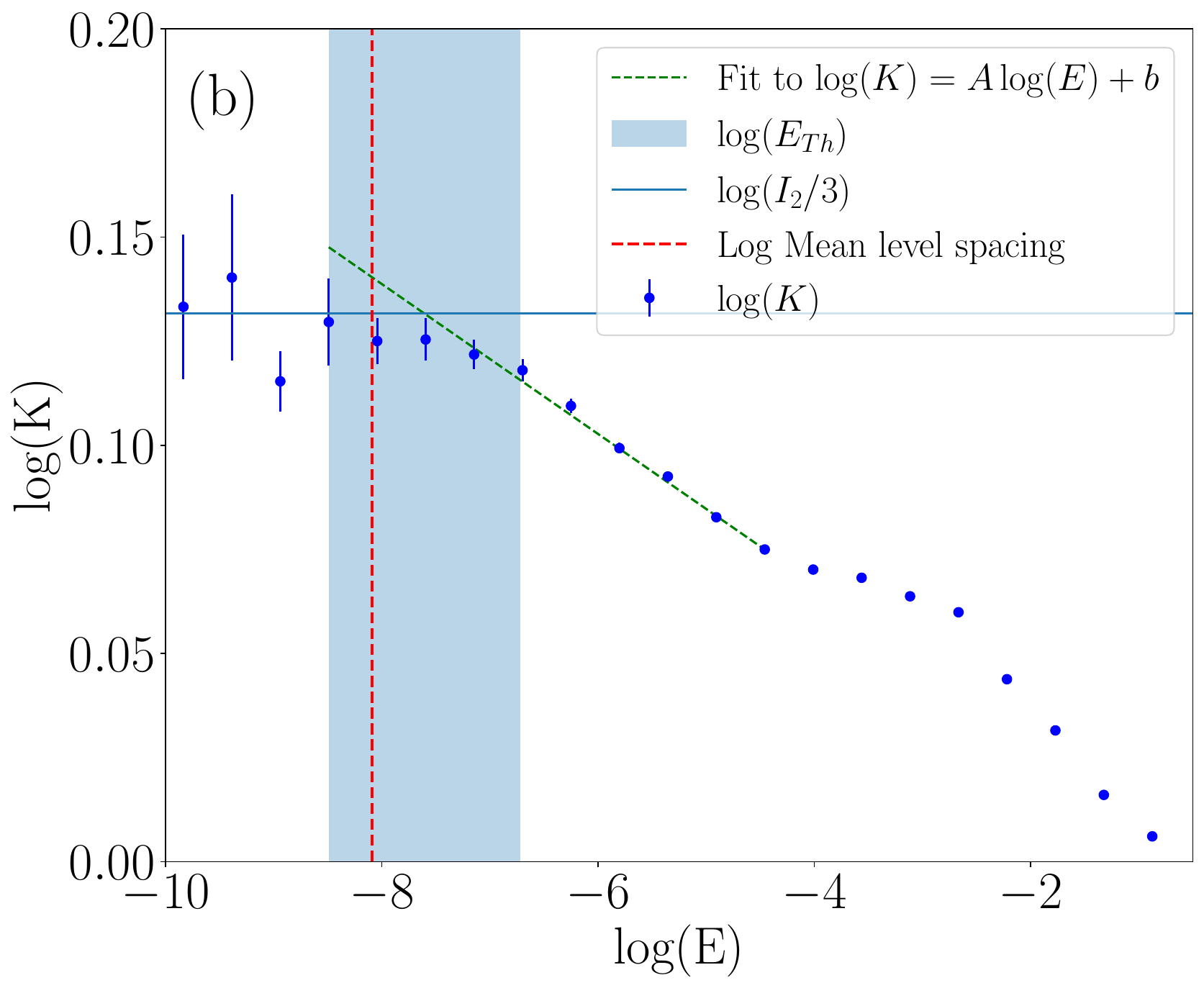}
    \caption{\label{fig:K}
Eigenvector correlations as a function of energy for TBG. (a) Circular flake with $N = 30720$ and incommensurate twist angle $\theta =21^\circ$. (b) Commensurate supercell flake with $\theta_{com}= 21.8^\circ$ and $N=31024.$ Scatter points indicate the eigenvector correlations $K$ averaged over many states around energy $4$ eV, while the horizontal blue line is the average value of $I_2/3.$ The blue-colored area is our estimated range of the Thouless energy, while the red dashed line corresponds to the mean level spacing. }
    \label{fig:dos_figuras}
\end{figure}

There are several ways to compute the Thouless energy. The first one is to use the rigidity of the eigenstates under a twist in the boundary conditions\ \cite{edwards_numerical_1972, anderson_thouless_1980, abrahams_scaling_1979}. However, this is not possible due to our choice of open boundary conditions.
An alternative computational tool for determining the Thouless energy, which does not rely on the previously discussed twist, is based on the density-density correlations between eigenstates and can be applied to any system\ \cite{kravtsov1997,kravtsov_random_2015, pino_multifractal_2017}, even many-body ones. We therefore resort to this numerical approach to compute $E_{Th}$. 
We first numerically determine the density-density correlator at energy $E$\ \cite{PhysRevB.96.214205}
\begin{equation}\label{eq:K}
     K(E)=\sum_{\alpha>\beta,i} \delta(E-E_\alpha+E_\beta)|\psi_{\alpha}(i)|^2|\psi_{\beta}(i)|^2,
\end{equation}
where $\alpha$ and $\beta$ label the eigenenergy levels and $i$ runs over the atomic positions. In our case, the sum will run over $\alpha,\ \beta$ for many states within an energy window centered at $E=4$ eV, the same energy window for which we computed the $r$ statistics in previous Section III. We can then determine the Thouless energy as the energy for which the structure of eigenstates follows the prediction of the relevant Random Matrix Ensemble, which in our case is the GOE \cite{mehta2004random}. Thus, we can determine $E_{Th}$ as the energy for which the correlator $K$ reaches the GOE prediction $K(E_{Th})=I_2/3$. Here, $I_2=\sum_i |\psi_i|^4$ is the second moment of the wavefunction\ \cite{pino_multifractal_2017}. 

In Fig.\ \ref{fig:K},  we present the $K$-correlator given by Eq.\ \ref{eq:K} for two different TBG lattices. Panel (a) corresponds to an incommensurate circular flake with  $N=30720$ sites and $\theta = 21^\circ,$  while panel (b) showcases a commensurate supercell flake with angle $\theta_{com}=21.8^\circ$ and $N=31024$. We aim for our commensurate and incommensurate lattices to be as similar as possible in terms of number of atoms and rotation angle. 
As already discussed at the end of Section III, in this section we will focus on larger angle TBG flakes.

Back to Fig.\ \ref{fig:K}, we find that for both panels (a) and (b)
the $K$-correlations 
exhibit a GOE plateau at small energies, in agreement with our previous discussion 
on the GOE prediction $K(E_{Th})=I_2/3$. 
We set $E_{Th}$ as the point in which this plateau is reached, and the exact value is decided by means of two numerical criteria. 
In the first method, dubbed $E_{Th}^{(1)},$ we choose the maximum energy at which the $K$-correlator 
reaches the GOE prediction. For the second one, labeled $E_{Th}^{(2)},$ we fit several points with energy larger than $E_{Th}^{(1)}$ to a linear function (green dotted lines in Fig.\ \ref{fig:K}) and define $E_{Th}$ as the point at which this line reaches the GOE plateau. We finally estimate the logarithm of the Thouless energy as $\log(E_{Th})= \log(E_{Th}^{(1)})$ 
together with its uncertainty as $\Delta \log(E_{Th})=\left|\log\left(E_{Th}^{(1)}\right)-\log\left(E_{Th}^{(2)}\right)\right|.$

\begin{figure}[t!]
    \centering
    \includegraphics[width=\columnwidth]{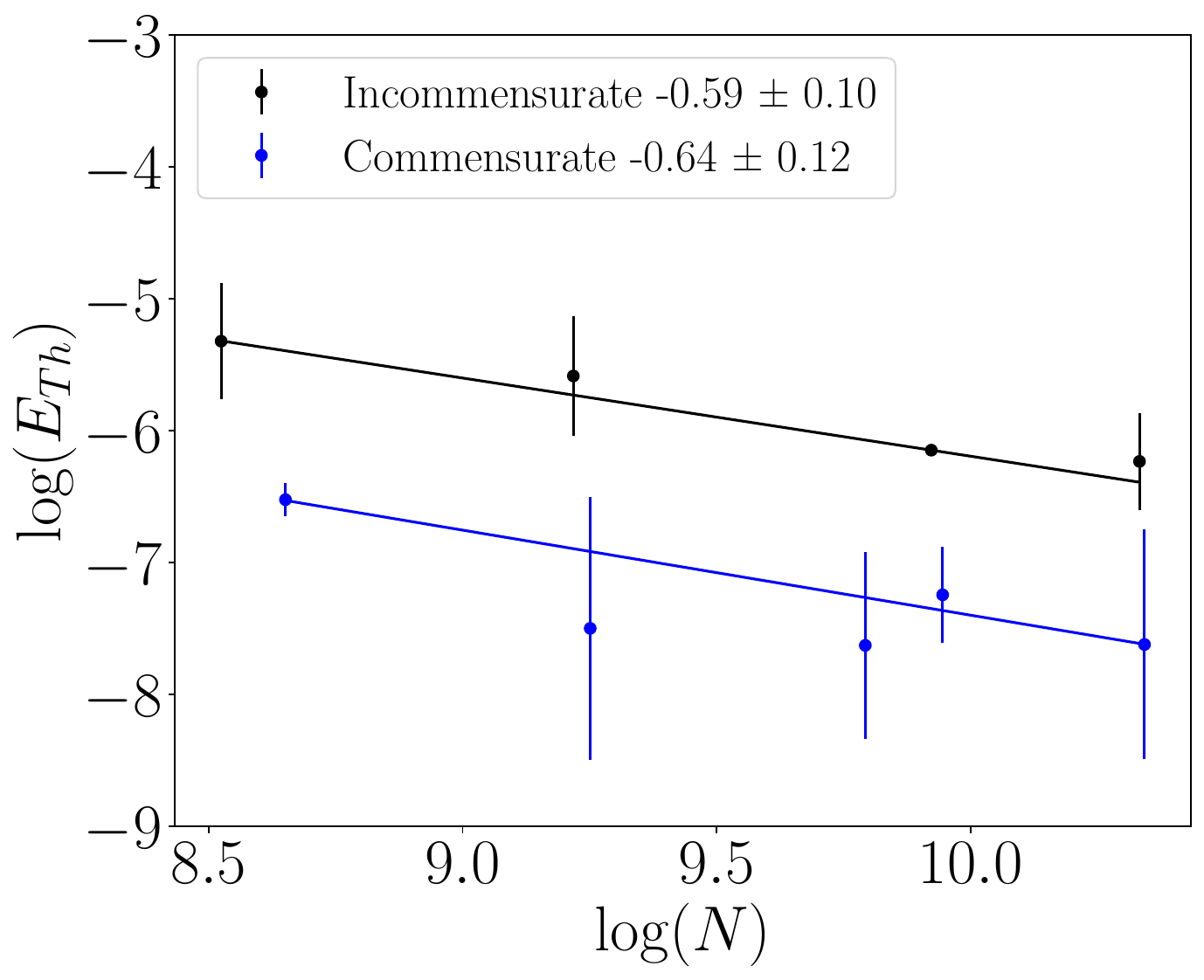}
    \vspace{0.5cm}
    \includegraphics[width=\columnwidth]{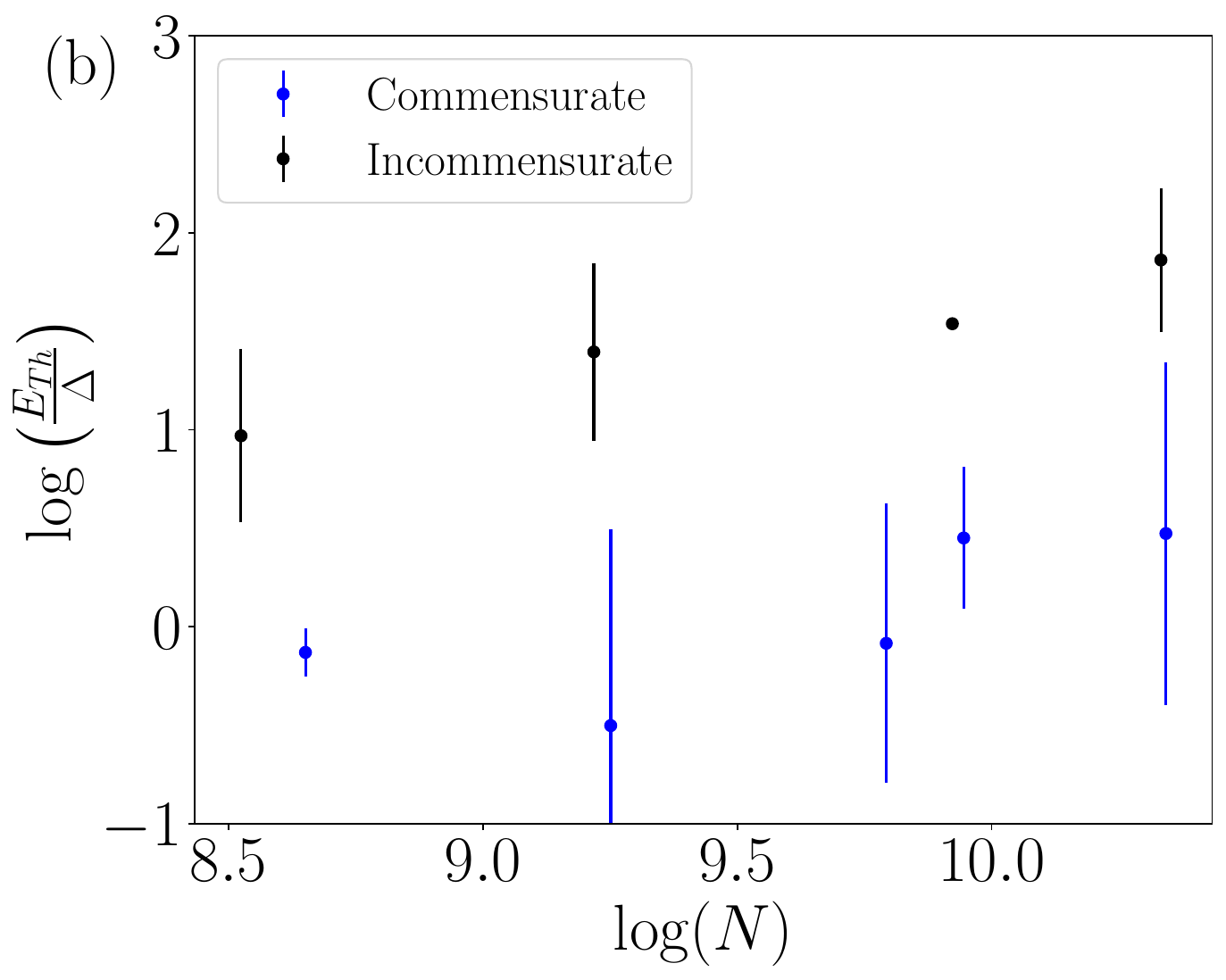}
    \caption{\label{fig:scaling}
Scaling of the Thouless energy $E_{Th}$ for the same commensurate (blue) and incommensurate (black) flakes as in Fig. \ref{fig:K}, with rotation angles around $21^\circ$ and $N\sim31000$ atoms. 
(a) Logarithm of the Thouless energy as a function of the logarithm of the system size. The solid lines depict fits of each data set to a linear function, which results in slopes $-0.59\pm0.10$ and $-0.64\pm 0.12$ for the incommensurate and commensurate structures, respectively. 
(b) Logarithm of the dimensionless conductance $\log\left(E_{Th}/\Delta\right)$ as a function of $\log(N)$ for the same flakes as in panel (a).}
\end{figure}

On a different note, we can also highlight the relation between our Thouless energy estimations and the dimensionless conductance. The dimensionless conductance is defined as $g = E_{Th}/\Delta,$ with $\Delta$ being the mean level spacing (red dotted lines in Fig.\ \ref{fig:K}). 
This quantity characterizes whether the system shows metallic or insulating  behavior \cite{abrahams_scaling_1979}. By comparing panels (a) and (b) of Fig.\ \ref{fig:K}, we can already notice that the dimensionless conductance is approximately $g\approx 1$ for the commensurate structure, while is $g> 1$ for the incommensurate case. At a qualitative level, we therefore find a more robust metallic behavior for the incommensurate lattice than for the commensurate system. We will further comment on the dimensionless conductance at the end of the following Section IV B.

\subsection{Scaling of the Thouless energy}

Once that we have discussed our procedure to compute the Thouless energy and estimated its value for commensurate and incommensurate lattices, we can study its scaling with the system size. As explained at the beginning of Section IV, Eq.\ \ref{Eq:Eth} implies that any intrinsic mechanism of decay would produce a mean free path $E_{Th}\sim 1/N$ (with $N$ being the number of atoms), while a border-induced diffusion would mean a size-dependent mean free path with $E_{Th}\sim 1/\sqrt{N},$ where we have used $L\sim \sqrt{N}.$  

Fig.\ \ref{fig:scaling} (a) depicts the logarithm of the Thouless energy as a function of the logarithm of the number of atoms for the same commensurate and incommensurate systems as depicted in Fig.\ \ref{fig:dos_figuras}. As discussed in the previous Section IV A, these systems host rotation angles $\theta_{com}=21.8^\circ$ and $\theta=21^\circ$, and are shown in Fig.\ \ref{fig:scaling} (a) in blue and black colors, respectively. We have fitted each data set to a law $\log(E_{Th})= a \log(N)+b,$ finding $a_{com}=-0.64\pm0.12$ and $a_{incom}=-0.59\pm0.10$. 
For both systems, a slope of -0.5 is compatible with our fitted slopes, indicating a scaling $E_{Th}\sim 1/\sqrt{N}$ and, as discussed before, a mean free path $l \sim L$ (so that $E_{Th}\sim 1/L$ using $L\sim \sqrt{N}$). Fig.\ \ref{fig:scaling} (a) therefore shows that the chaotic behavior of both commensurate and incommensurate large angle TBG flakes is caused by borders. 
We should however note that the linear fit for the incommensurate case is more reliable than the commensurate one, as error bars 
are larger for the latter (in log-scale). This is in agreement with Fig. \ref{fig:rs_angle_p(r)} of Section III, and overall with our previous discussions regarding the crucial role of borders in the chaotic spectra presented in this work. We confirm that commensurability plays a secondary role for the flake geometries and energy windows we have worked with.

There is, however, an interesting difference between our commensurate and incommensurate flakes. From Fig. \ref{fig:K}, we note that the Thouless energy is significantly smaller for the commensurate flake. As discussed in Section IV A, these differences in the Thouless energy also imply notably different values of the dimensionless conductance $g=E_{Th}/\Delta$. 
To further analyze this, in Fig.\ \ref{fig:scaling} (b) we plot the dimensionless conductance  
as a function of the logarithm of the system size $N$. Although the error bars are large for both commensurate and incommensurate flakes, we confirm our discussion of Section IV A: for the incommensurate case $g$ is compatible with a normal metallic behavior, and the Thouless energy is significantly larger than the mean level spacing. The commensurate lattice, however, 
displays a small dimensionless conductance close to one.

\section{Conclusions}

We have analyzed the emergence of chaos and diffusion
in finite-size TBG systems, spanning commensurate and incommensurate lattices as well as various rotation angles, with an emphasis on larger angles. We study their energy spectra and develop a new framework for evaluating the source of chaos in these lattices, based on the Thouless energy.

Our results show that chaos can be easily induced irrespectively of whether the lattice is commensurate or incommensurate, although the primary mechanism that induces it can be different depending on the rotation angle for commensurate lattices. For small rotation angles, chaos can appear even in commensurate supercells if the size of the unit cell is comparable to the system size \cite{porter_chaos_2017}.
At larger angles, the commensurate unit cell is usually smaller and chaos mainly appears due to border-induced scattering. We have verified this hypothesis by computing the Thouless energy of both commensurate and incommensurate large angle $\theta \sim 21^\circ$ TBG flakes. The resulting Thouless energy is compatible with a mean free path of the order of the linear system size, characteristic of systems with boundary-induced chaos. This behavior is reported in both commensurate and incommensurate lattices.

However, we have found intriguing differences between commensurate and incommensurate systems in regard to the actual value of the Thouless energy. Commensurate flakes showcase a small Thouless energy, similar to the mean level spacing, and a related small dimensionless conductance. Since our TBG commensurate supercells have the shape of 
a regular rhombus, this difference may be related to the non-integrable (but neither chaotic) character of a quantum billiard with that shape, as discussed in Refs.\ \cite{Bogomolnymodels1999,gremaud1998spacing}. 

Our results bear implications for transport measurements in TBG. Indeed, a good understanding of the main scattering process affecting conductivity is key\ \cite{delgado-notario_unveiling_2025,wang_one-dimensional_2013}. Experiments with nanoconstrictions in TBG, as in Ref.\ \cite{clerico_quantum_2019} where ballistic behavior arises in graphene, may be much more difficult to achieve in TBG due to the border rugosity. On the theoretical side, our work demonstrates that TBG behaves in a more complex manner when boundaries are taken into account, which may affect a putative Anderson localization transition as the one described in\ \cite{goncalves_critical_2023, goncalves_renormalization_2023}. It has also been shown that changing the type of edges can lead a transition from Poisson-like to GOE statistics in monolayer graphene\ \cite{yu_gaussian_2016}. 

Moreover, our Thouless energy analysis could provide further information regarding the mean free path due to incommensurability in TBG.  A crossover from $E_{Th}\sim 1/L$ to $E_{Th}\sim 1/L^2$ would be expected for system sizes larger than the mean free path. This could provide valuable information on the intrinsic scattering mechanisms in TBG, or other moiré systems,  at different angles. We plan to improve our numerical techniques using polynomial filter diagonalization techniques\ \cite{pino_correlated_2024} to reach larger system sizes in future works. 

In summary, by combining well-known tools such as the adjacent gap ratios and probability distributions with new Thouless energy estimations, we have performed a detailed study of the chaotic dynamics of TBG flakes beyond magic angle physics.  

\section{Acknowledgments}

We thank J. Chalker, A. Rodríguez and J. Caridad for helpful discussions. This work is part of the European Union NextGeneration EU/PRTR project Consolidación Investigadora CNS2022-136025. M. P. acknowledges further support through grant no. PID2024-156340NB-I00 funded by Ministerio de Ciencia, Innovación y Universidades/Agencia Estatal de Investigación (MICIU/AEI/10.13039/501100011033) and the European Regional Development Fund (ERDF).  The numerical computations were performed in the facilities of Supercomputación Castilla y León (SCAYLE).

\let\selectlanguage\relax 
\bibliographystyle{apsrev4-2}
\bibliography{ref}

\end{document}